\newcommand{\Tr}{\mathop{\mathrm{Tr}}\nolimits}

\documentclass[pra,twocolumn,preprintnumbers,showpacs,eqsecnum]{revtex4}

\usepackage{amsfonts,amssymb,bm,graphicx}

\begin{document}

\title[Phase in finite-dimensional quantum systems]
{A complementarity-based approach to
phase in finite-dimensional quantum systems}

\author{Andrei B. Klimov}
\affiliation{Departamento de F\'{\i}sica,
Universidad de Guadalajara, Revoluci\'on~1500,
44420~Guadalajara, Jalisco, Mexico}

\author{Luis L. S\'{a}nchez-Soto}
\affiliation{Departamento de \'Optica,
Facultad de F\'{\i}sica, Universidad
Complutense, 28040 Madrid, Spain}

\author{Hubert de Guise}
\affiliation{Department of Physics,
Lakehead University, Thunder Bay,
Ontario P7B 5E1, Canada}

\date{\today}

\begin{abstract}
We develop a comprehensive theory of phase
for finite-dimensional quantum systems.
The only physical requirement we impose
is that phase is complementary to
amplitude. To implement this complementarity
we use the notion of mutually unbiased bases,
which exist for dimensions that are powers of
a prime. For a $d$-dimensional system (qudit)
we explicitly construct $d+1$ classes of
maximally commuting operators, each one
consisting of $d-1$ operators. One of this
class consists of diagonal operators that
represent amplitudes (or inversions). By
the finite Fourier transform, it is
mapped onto ladder operators that can
be appropriately interpreted as phase
variables. We discuss the examples of
qubits and qutrits, and show how these
results generalize previous approaches.
\end{abstract}
\pacs{03.65.Ta, 03.65.Ca, 03.65.Ud}

\maketitle

\section{Introduction}

The standard formalism of quantum optics is
usually presented in the context of the harmonic
oscillator, where both position and momentum
are represented by unbounded operators with
eigenvalues in the real numbers. Systems
living in a finite-dimensional Hilbert space
were studied originally by Weyl~\cite{Weyl50}
and also by Schwinger~\cite{Schwinger60},
but except for some relevant exceptions
(for a complete review see~\cite{Vourdas04}),
they have received the attention they rightly
deserve only after becoming one essential ingredient
in the development of the emerging field of
quantum information~\cite{Nielsen01,Zeilinger00}.
Indeed, the promise of futuristic technologies
like safe cryptography and new  ``super computers",
capable of handling otherwise untractable problems,
relies on the ability to control the quantum states of
a small number of qubits~\cite{Galindo02,Keyl02}.

In the modern parlance of quantum information
the concept of phase for a $d$-dimensional
system (or qudit) is ubiquitous. However,
in spite of being a primitive of the
theory, this notion is rather imprecise and,
roughly speaking, three quite distinct
conceptions can be discerned.

In the first, phase is considered as a
parameter and the problem is reduced to
the optimal estimation of the value of the
phase shift undergone by the qudit under
certain operations~\cite{phaseest}. Although
very operational in style, it accommodates
perfectly the practical requirements of
typical applications handled in quantum
information.

In the second, a semiclassical approach is
adopted, and the phase is assumed to be
linked to the geometry of the state
space: for example, for a qubit this
space is the well-known Poincar\'e sphere and
the phase is identified with the angle that
a state representative makes with the $Z$
axis~\cite{phaseparam}. This pictorial
understanding of phase as an angle makes
easy contact with the classical world,
but once more considers the phase as a
mere state parameter instead of a full
quantum variable.

The third major concept emphasizes the idea
that phase is a physical property and, by any
orthodox picture of quantum mechanics, must
be associated with a selfadjoint operator (or at
least with a family of positive operator-valued
measures). In this vein, phase operators
have been constructed via a polar decomposition
for qubits and qutrits~\cite{phaseop}.

The main goal of this paper is to look at the
fundamental problem of properly defining phase
from quite a different perspective. On closer
examination, one immediately discovers
that the idea of complementarity is at the
root of all the previous approaches: phase
is complementary to some amplitude, by
which we loosely mean that the precise
knowledge of one implies that all possible
outcomes of the other are equally
probable~\cite{Wheeler83}. This idea of
\textit{unbiasedness} leads directly
to introduce mutually unbiased bases
(MUBs)~\cite{Wootters87}, which, for a
variety of reasons, are becoming an
important tool in quantum optics~\cite{MUBs}.
It is known that the  maximum number of such
bases cannot be greater than $d +1$ and that
this limit is certainly reached if $d$ is a
power of a prime~\cite{Ivanovic81}. It is not
known if there are nonprime-power values of
$d$ for which this bound is attained. We
shall be not concerned with this problem
in this paper, and assume that we are
always working in a prime dimension.

It is essential to recall that complementarity
for the position-momentum pair is implemented
by the Fourier transform, which exchanges both
operators. Using the ideas introduced in
Ref.~\cite{Bandyopadhay02}, we construct
$d+1$ disjoint classes of maximally commuting
unitary matrices (each set having $d-1$ operators).
We then note that one of these classes
consists solely of diagonal operators
(which we can relate to \textit{inversions})
that can be mapped, using the finite Fourier
transform, to operators acting cyclically
on basis states.

This perspective leads to a natural notion of
\textit{phases} as complementary to
\textit{inversions}. The advantage of this
approach is that it does not rely on polar
decompositions or semiclassical arguments
and provides a clear understanding of the
behavior of such basic variables.

\section{Multicomplementary operators for
finite-dimensional quantum sytems}

The objects we study in this paper are quantum
systems described in a $d$-dimensional Hilbert
space $\mathcal{H}_d$. We recall~\cite{Wootters87}
that two different orthonormal bases $\mathcal{A}$
and $\mathcal{B}$ are said to be mutually
unbiased if a system prepared in any element
of $\mathcal{A}$ (such as $| a \rangle$) has
a uniform probability distribution of being found
in any element of $\mathcal{B}$
\begin{equation}
\label{unbias}
|\langle a | b \rangle |^2 = \frac{1}{d} ,
\end{equation}
for all $a \in \mathcal{A}$ and all $b \in
\mathcal{ B}$. As anticipated in the Introduction,
we shall be concerned solely in cases where $d$
is a prime number, as then we know that there
are $d+1$ MUBs. For dimensions which are power
of a prime the argument can be easily extended
with some modifications~\cite{fields05}

If $| n \rangle $ ($n = 0, \ldots, d-1$) is the
standard (computational) basis in $\mathcal{H}_d$,
we introduce the generalized Pauli matrices $X$ and
$Z$ by the following action:
\begin{eqnarray}
X | n \rangle & = & | n + 1 \rangle  ,
\nonumber \\
& & \\
Z | n \rangle & = & \omega^n | n \rangle ,
\nonumber
\end{eqnarray}
where
\begin{equation}
\omega = \exp (2 \pi i /d) .
\end{equation}
Note that throughout this paper addition and
multiplication must be understood $\bmod \ d$.
These operators $X$ and $Z$, which are
generalizations of the Pauli matrices, were
studied by Patera and Zassenhaus~\cite{Patera88}
in a purely mathematical context, and
have been used recently by many authors
in a variety of applications~\cite{Gottesman01}.
Under multiplication, they generate a finite subgroup
of SU($d$), known as the generalized Pauli group,
and obey the finite-dimensional version of the
Weyl form of the commutation relations:
\begin{equation}
\label{XZo}
Z X = \omega X Z .
\end{equation}
It is easily shown that the eigenvectors of $X$
and those of  $Z$ satisfy (\ref{unbias}).

To simplify as much as possible the following
computation, we introduce the following labeling
scheme: let
\begin{equation}
\mathfrak{X}_0 = Z,
\qquad
\mathfrak{X}_k = X Z^{k-1} ,
\qquad
k = 1, \ldots, d .
\end{equation}
Since we shall also need powers of these
operators, we denote by
$\mathfrak{C}_{\mathfrak{X}_k}$ the
set
\begin{equation}
\label{class}
\mathfrak{C}_{\mathfrak{X}_k}= \{ \mathfrak{X}_k,
\mathfrak{X}_k^2, \ldots,   \mathfrak{X}_k^{d-1} \} .
\end{equation}
The $d-1$ operators in the class
$\mathfrak{C}_{\mathfrak{X}_k}$ clearly
commute one with another and therefore
represent a maximal set of commuting operators.

Following the ideas in Ref.~\cite{Bandyopadhay02},
consider now the following set (each containing
$d+1$ operators)
\begin{eqnarray}
\label{set}
\mathcal{S} & = &
\{ \mathfrak{X}_0, \mathfrak{X}_1, \ldots,
\mathfrak{X}_d \} \nonumber \\
&  = &  \{Z, X, XZ, \ldots, XZ^{d-1} \} .
\end{eqnarray}
By virtue of the relation (\ref{XZo}),
any two operators in this set are
complementary, in the sense that
their eigenvectors satisfy the
unbiasedness condition (\ref{unbias}).
Furthermore, a complete MUB is obtained
by constructing every eigenvector of every
element in $\mathcal{S}$, so we refer
to $\mathcal{S}$ as a maximal set
of multicomplementary operators.

In the case of the standard position-momentum
complementary variables, their eigenvectors form
bases related by the Fourier transform. The
finite-dimensional Fourier transform can
be defined as~\cite{Ip02}
\begin{equation}
F= \frac{1}{\sqrt{d}} \sum_{n,n^\prime=0}^{d-1}
\omega^{n n^\prime}|n \rangle \langle n^\prime | ,
\end{equation}
with the properties
\begin{equation}
F F^\dagger = F^\dagger F = 1 ,
\qquad
F^4 = 1 .
\end{equation}
Using this definition one can check that $X$ and $Z$
are indeed Fourier pairs
\begin{equation}
X = F^\dagger Z F .
\end{equation}
There exist also an operator $V$ that transform
$X \rightarrow X Z^k$. It has a diagonal form:
\begin{equation}
\label{Vodd}
V = \sum_{n=0}^{d-1} \omega^{-  (n^2 - n ) (d+1)/2 }
|n \rangle \langle n| ,
\end{equation}
so that
\begin{equation}
\label{Vdiag}
X Z^k = V^\dagger{}^k  X V^k ,
\end{equation}
where we have assumed an odd-prime dimension (the
case $d=2$ need minor modifications, as we shall
see in next Section).

In physical applications, only $d-1$ populations
can vary independently in a $d$-level system.
In consequence, it is usual to work with
$d-1$ traceless operators $h_j$ that
measure population inversions between
the corresponding levels, i.e.,
\begin{equation}
h_j = S_{jj} - S_{j+1 j+1} ,
\end{equation}
where
\begin{equation}
\label{Sij}
S_{ij} = |i \rangle \langle j | .
\end{equation}
These $h_j$ (usually known as the
Cartan-Weyl generators) constitute
a maximal Abelian subalgebra. Note that the
diagonal operators in the class
$\mathfrak{C}_{\mathfrak{X}_0}  =
\{ Z^k \}$ are linear combinations of
$h_j$, so both can be used indistinctly.

On physical grounds, we expect phases
to be complementary to inversions.
But inversions are invariant under
phase shifts: if
\begin{equation}
\label{U}
U (\bm{\varphi}) = \exp \left ( -
i \sum_j \varphi_j h_j \right ) ,
\end{equation}
where $\bm{\varphi}$ denotes
$(\varphi_1, \ldots, \varphi_{d-1})$,
then
\begin{equation}
U^\dagger (\bm{\varphi}) \, h_k \,
U (\bm{\varphi} ) = h_k .
\end{equation}
Thus, we can construct a continuous family
of operators, all complementary to inversions,
by conjugating any operator in the class
$\mathfrak{C}_{\mathfrak{X}_k}$ by
$U (\bm{\varphi})$. In particular,
the diagonal operator $V$, which maps $X$ to
$XZ^k$ as per Eq.~(\ref{Vdiag}) (and that is
also of the form (\ref{U}) for a definite
choice of the parameters), allows us to
construct any $\mathfrak{X}_k$ starting
with $\mathfrak{X}_1= X$. For this reason,
and without any loss of generality, we
shall henceforth focus on the elements of
the class $\mathfrak{C}_{\mathfrak{X}_1} =
\{ X, X^2, \ldots, X^{d-1}\}$ to
discuss general properties of complementary
phase operators. We thus define $d-1$
families of  operators representing the
exponential of the phase by
\begin{equation}
\label{Ek}
E^k (\bm{\varphi}) =
U^\dagger (\bm{\varphi}) \, X^k
\, U (\bm{\varphi}) ,
\qquad k = 1, \ldots, d-1,
\end{equation}
obtained by successive powers.

If $| s \rangle$ is an eigenstate of
$h_j$ with eigenvalue $h_{js}$, then the
expectation value of $E^k (\bm{\varphi})$
on an arbitrary state $|\psi \rangle =
\sum_s c_s |s \rangle$ is simply
\begin{equation}
\langle E^k (\bm{\varphi} ) \rangle =
\sum_{s} c_{s+k}^\ast c_s \,
U_{s+k}^\ast (\bm{\varphi})
U_s (\bm{\varphi}) ,
\label{Emed}
\end{equation}
where
\begin{equation}
U_s (\bm{\varphi}) = \exp \left ( -
i \sum_j \varphi_j h_{js} \right ) .
\end{equation}
This allows us to introduce an operator
kernel
\begin{equation}
\Pi ( \bm{\varphi} )  =
\frac{1}{( 2 \pi )^{d-1}}
\left [ \openone + \sum_{k=1}^{d-1}
\langle E^{k} (\bm{\varphi}) \rangle^\ast \,
X^{k} \right ] ,
\end{equation}
which is properly normalized and generates
all the moments through the relation
\begin{equation}
\langle E^{l} (\bm{\varphi}) \rangle =
\frac{ ( 2\pi )^{d-1}}{d}
\Tr [ \Pi ( \bm{\varphi} ) X^{l} ] ,
\end{equation}
so the phase distribution is obtained as
\begin{equation}
P ( \bm{\varphi} ) =
\frac{1}{( 2 \pi )^{d-1}}
\left( 1 + \sum_{k=1}^{d-1}
\langle E^{k} (\bm{\varphi} ) \rangle ^\ast
\, \langle X^{k} \rangle \right ) .
\label{Pdf}
\end{equation}
We note in passing that any positive
operator-valued measure of the
general form
\begin{equation}
\label{POM}
\Delta ( \bm{\varphi} )  =
\frac{1}{( 2 \pi )^{d-1}}
\left [ \openone + \sum_{k=1}^{d-1}
\gamma_k \, E^{k} (\bm{\varphi}) \right ] ,
\end{equation}
which can be associated to (\ref{Pdf}),
satisfies the usual requirements of real
valuedness, positivity and normalization
and  possesses the obvious property
\begin{equation}
e^{i \varphi_j^\prime h_j}
\Delta (\varphi_1, \ldots, \varphi_{d-1})
e^{-i \varphi_j^\prime h_j} =
\Delta (\varphi_1, \ldots, \varphi_j +
\varphi_j^\prime, \ldots, \varphi_{d-1}),
\end{equation}
which meets the usual requirements of
complementarity~\cite{compovm}.

\section{Application: quantum phase for
finite-dimensional systems}

\subsection{The case of qubits}

To fully appreciate the details of the method,
we shall work out some relevant examples. First,
we focus on the simplest case of a two-dimensional
Hilbert space $\mathcal{H}_2$, and a state space
that coincides with the sphere $S_2$.

The basic operators are the standard Pauli matrices
\begin{equation}
X = 2 \sigma_x = \left (
\begin{array}{rr}
0 & 1 \\
1 & 0
\end{array}
\right ) ,
\qquad
Z = 2 \sigma_z =
\left (
\begin{array}{rr}
1 & 0 \\
0 & -1
\end{array}
\right )  ,
\end{equation}
such that
\begin{equation}
\sigma_x \sigma_z = - \sigma_z \sigma_x .
\end{equation}
The  transformation $\sigma_z \rightarrow \sigma_x$
is accomplished by the finite Fourier transform
\begin{equation}
\label{F2}
F=\frac{1}{\sqrt{2}}
\left (
\begin{array}{rr}
1 & 1 \\
1 & -1
\end{array}
\right )  .
\end{equation}
However, it is impossible to find a unitary
transformation $V$ such that $\sigma_x \rightarrow
\sigma_x \sigma_z$. For this reason, instead
of $\sigma_x \sigma_z$ the matrix $\sigma_y =
i \sigma_x \sigma_z$ is used, so that
$\sigma_y =  V^\dagger \sigma_x V$,
where $V$ is the unitary operator
\begin{equation}
V=
\left(
\begin{array}{rr}
1 & 0 \\
0 & -i
\end{array}
\right ) .
\end{equation}

If  $\varrho_A$ describes a point in the
Bloch sphere $S_2$ in an arbitrary direction
parametrized by $\mathbf{n} =
(\cos \varphi_A \sin \vartheta_A ,
\sin \varphi_A \sin \vartheta_A ,
\cos \vartheta_A)$, let $A$ be
\begin{equation}
\label{su2lc1}
A = \mathbf{n}_A \cdot \bm{\sigma} =
R ( \vartheta_A ,\varphi_A ) \ \sigma_z \
R^{-1} ( \vartheta_A ,\varphi_A ) \, ,
\end{equation}
where
\begin{equation}
R ( \vartheta , \varphi ) = \exp \left [
\frac{\vartheta}{2} \left (
\cos \varphi \ \sigma_x   -
\sin \varphi \ \sigma_y
\right) \right ] \,  .
\end{equation}
The condition of complementarity between $A$
and a generic operator $B$ [expressed also
as in Eq.~(\ref{su2lc1})] can be written
as~\cite{Kim03}
\begin{equation}
\label{comBloch}
\mathbf{n}_A \cdot \mathbf{n}_B = 0 \, ;
\end{equation}
that is, the subspace spanned in $S_2$ by
$\mathbf{n}_A$ is orthogonal to that by
$\mathbf{n}_B$. We have then a one-parameter
set of complementary operators of the general
form
\begin{equation}
B = \mathbf{n}_B \cdot \bm{\sigma} \, ,
\end{equation}
where the unit vector $\mathbf{n}_B$ satisfies
(\ref{comBloch}), which is equivalent to
\begin{equation}
\cot \vartheta_B = - \tan \vartheta_{A}
\cos ( \varphi_B - \varphi_{A} ) \, .
\end{equation}
In particular, the complementary set to
the inversion $\sigma_{z}$ consists in
the one-parameter family
\begin{equation}
\label{E}
E (\varphi) = \cos \varphi  \ \sigma_x -
\sin \varphi \ \sigma_y
=
\left (
\begin{array}{cc}
0 & e^{i \varphi} \\
e^{- i \varphi} & 0
\end{array}
\right ) ,
\end{equation}
where $\varphi$ represents a reference phase.
This in fact agrees with the exponential of the
phase operator obtained via a polar
decomposition~\cite{phaseop}.

According to the approach developed in this
paper, we have now one family of phase operators
that can be constructed as
\begin{equation}
E (\varphi) =   \exp ( i\varphi \sigma_z/2) \,
\sigma_x \,
\exp (-i\varphi \sigma_z/2) .
\end{equation}
This coincides also with (\ref{E}), and can
be recast in the suggestive form
\begin{equation}
\label{E2}
E (\varphi) = \underline{e} ( \varphi ) {}^t
F  \underline{\sigma} ,
\end{equation}
where $t$ denotes the transpose and
\begin{equation}
\underline{e} (\varphi) = \frac{1}{\sqrt{2}}
\left (
\begin{array}{c}
e^{- i \varphi } \\
e^{i \varphi }
\end{array}
\right ) ,
\qquad
\underline{\sigma} =
\left (
\begin{array}{c}
\sigma_x \\
\sigma_x \sigma_z
\end{array}
\right)  .
\end{equation}
This result confirms in this simple case the
complementary character of $E(\varphi) $ obtained
via Fourier transform. For a pure state
such as
\begin{equation}
| \psi \rangle =
\left (
\begin{array}{c}
\cos (\vartheta/2) \\
\sin (\vartheta/2) \, e^{i \chi }
\end{array}
\right ) ,
\end{equation}
with $0 \le \vartheta \le \pi$,
$0 \le \chi \le 2 \pi$, the average
value of $E(\varphi)$ is
\begin{equation}
\langle E (\varphi) \rangle =
\sin \vartheta
\cos (\chi + \varphi ) .
\end{equation}
The main features of this description
are obviously independent of the reference
phase $\varphi$.

\subsection{The case of qutrits}

For a three-dimensional  Hilbert space
$\mathcal{H}_3$ the basic operators are
\begin{equation}
X =
\left (
\begin{array}{ccc}
0 & 0 & 1 \\
1 & 0 & 0 \\
0 & 1 & 0
\end{array}
\right ) ,
\qquad
Z =
\left (
\begin{array}{ccc}
1 & 0 & 0 \\
0 & \omega & 0 \\
0 & 0 & \omega^2
\end{array}
\right ) ,
\end{equation}
and  $\omega = \exp (2 \pi i/3)$.
We have four classes of disjoint traceless
operators, each containing two commuting
operators
\begin{eqnarray}
& \mathfrak{C}_{\mathfrak{X}_0} = \{ Z, Z^2\}  ,
\quad
\mathfrak{C}_{\mathfrak{X}_1} = \{ X, X^2\} , &
\nonumber \\
& & \\
& \mathfrak{C}_{\mathfrak{X}_1} =
\{X Z, (X Z)^2\},
\qquad
\mathfrak{C}_{\mathfrak{X}_3} =
\{X Z^2, (X Z^2)^2 \} . &
\nonumber
\end{eqnarray}

The discrete Fourier transform is
\begin{equation}
F=\frac{1}{\sqrt{3}}
\left (
\begin{array}{ccc}
1 & 1 & 1 \\
1 & \omega & \omega ^{2} \\
1 & \omega ^{2} & \omega
\end{array}
\right ) ,
\label{F3}
\end{equation}
and $V$ is the diagonal unitary matrix
\begin{equation}
V=
\left (
\begin{array}{ccc}
1 & 0 & 0 \\
0 & 1  & 0 \\
0 & 0 & \omega^2
\end{array}
\right ) .
\end{equation}

In this case we have two Cartan operators
associated with the two independent inversions
\begin{equation}
h_1 =
\left (
\begin{array}{rrr}
1 & 0 & 0 \\
0 & -1 & 0 \\
0 & 0 & 0
\end{array}
\right ) ,
\qquad
h_2 =
\left (
\begin{array}{rrr}
0 & 0 & 0 \\
0 & 1 & 0 \\
0 & 0 & -1
\end{array}
\right ) ,
\end{equation}
which can be easily expressed as linear combinations
of $Z$ and $Z^2$. Note that $X$ and $X^2$
correspond to two different physical situations:
in the computational basis $X$ acts as
 $X |n \rangle =|n + 1\rangle$, while
$X^2| n \rangle =|n + 2\rangle$.

Thus, we have two families of commuting phase
operators:
\begin{eqnarray}
\label{uff}
E (\varphi_1, \varphi_2)   & = &
e^{i(\varphi_1 h_1 + \varphi_2 h_2)} \, X \,
e^{-i(\varphi_1 h_1 + \varphi_2 h_2)} \nonumber \\
& & \\
E^2 (\varphi_1, \varphi_2) & = &
e^{i( \varphi_1 h_1 + \varphi_2 h_2)}
\, X^2 \, e^{-i( \varphi_1 h_1 + \varphi_2 h_2)}  .
\nonumber
\end{eqnarray}
In this particular case, they essentially
coincide because $E^2 (\varphi_1, \varphi_2) =
E^\dagger(\varphi_1, \varphi_2)$.  The
phase operator $E(\varphi_1, \varphi_2)$
can be represented in a form similar to
 (\ref{E2}), namely
\begin{equation}
E (\varphi_1, \varphi_2) =
\underline{e}(\varphi_1, \varphi_2){}^t   F
\underline{X} ,
\end{equation}
with
\begin{equation}
\underline{e}  (\varphi_1, \varphi_2) =
\frac{1}{\sqrt{3}}
\left (
\begin{array}{c}
e^{i (\varphi_2 - 2 \varphi_1 )} \\
e^{i(\varphi_1 + \varphi_2)} \\
e^{i(\varphi_1 - 2 \varphi_2)}
\end{array}
\right ) ,
\qquad
\underline{X}=
\left (
\begin{array}{c}
X \\
XZ \\
XZ^2
\end{array}
\right ) .
\end{equation}
Its explicit form is
\begin{equation}
E (\varphi_1, \varphi_2) =
\left (
\begin{array}{ccc}
0 & 0 &  e^{i ( \varphi_1 + \varphi_2 )} \\
e^{-i ( 2 \varphi_1 - \varphi_2 )} & 0 & 0 \\
0 & e^{i ( \varphi_1 - 2 \varphi_2 )} & 0
\end{array}
\right )  ,
\end{equation}
where again $\varphi_1$ and $\varphi_2$ are
reference phases. The average value of
$E(\varphi_1, \varphi_2)$ in an arbitrary
pure state such as
\begin{equation}
| \psi \rangle =
\left (
\begin{array}{c}
\cos (\vartheta /2)  \\
\sin (\vartheta /2)  \cos (\xi/2) \, e^{i \chi_1 } \\
\sin (\vartheta /2) \sin (\xi/2) \, e^{i \chi_2 }
\end{array}
\right ) ,
\end{equation}
is simply
\begin{eqnarray}
\langle E  (\varphi_1, \varphi_2) \rangle & = &
\frac{1}{2} \sin^2 (\vartheta/2)
\sin \xi e^{i[ (\chi_1 -\chi_2 ) +
(\varphi_1 - 2 \varphi_2 )]} \nonumber \\
& + & \frac{1}{2} \sin \vartheta
\{ \sin (\xi /2)  e^{i [\chi_2 + ( \varphi_1 + \varphi_2 )]}
 \nonumber \\
& + & \cos (\xi /2)  e^{-i[ \chi_1 + ( 2 \varphi_1 -
\varphi_2 )]}  \} .
\end{eqnarray}
We can observe that when only two levels are involved,
$\xi =0$,
\begin{equation}
\langle E (\varphi_1, \varphi_2) \rangle =
\frac{1}{2} \sin \vartheta e^{-i [\chi_1
+ ( 2\varphi_1 - \varphi_2 )] },
\end{equation}
which measure the relative phase between the states
1 and 2, and depends only on a single effective phase
$2 \varphi_1 - \varphi_2$. Since $E^2 (\varphi_1,
\varphi_2) = E^\dagger (\varphi_1, \varphi_2)$,
which includes all the possible moments,
the phase distribution (\ref{Pdf}) remains constant
at the direction $2 \varphi_1 - \varphi_2 =$
constant.

It is instructive to compare our construction
of phase operators with the more common
algorithm, based on the polar decomposition
of $S_{12}$, $S_{23}$, and $S_{13}$,
defined in Eq.~(\ref{Sij}).
The polar decomposition of $S_{ji}$ implicitly
focuses of the $i \rightarrow j$ transition,
without analyzing the role of the third
``spectator" state of the system. This
observation can be used to explain the
lack of uniqueness in the polar decomposition
of singular operators like $S_{12}$, for
instance. This approach produces a phase operator
$E_{12}$ of the form
\begin{equation}
\label{polar}
E_{12} =
\left (
\begin{array}{rrr}
0 & 1 & 0 \\
x & 0 & y \\
y^\ast & 0 & - x^\ast
\end{array}
\right ) ,
\qquad
|x|^2 + |y|^2 = 1,
\end{equation}
where $S_{12} = E_{12} R_{12}$ and
$R_{12} = \sqrt{S_{12} S_{21}}$ is the
``modulus". The only constraints on
$E_{12}$ are imposed by the requirement
of unitarity.

A particular solution to Eq.~(\ref{polar})
is obtained by ``isolating" the third state
from the first two by choosing $y=0$ and
thus $x = e^{i \varphi_{12}}$. Since
$(S_{12}, S_{21}, h_1)$ span a su(2)
subalgebra, this choice amounts to limiting
the action of the phase operator $E_{12}$
to a specified su(2) subspace. A similar
argument holds for $E_{23}$, with
$(S_{23}, S_{32}, h_2)$ spanning another
su(2) subalgebra.

This perspective in terms of polar
decomposition and transitions leads to
phase operators that fulfill the requirements
of complementarity only between pairs of
states involved in each transition.
The  corresponding positive operator-valued
measure obtained in this construction is
still of the general form
found in Eq.~(\ref{POM}). However, this
restricted point of vies is to be contrasted
with the approach of this paper, where
complementarity is imposed for the three-level
system as a whole.

\section{Concluding remarks}

Mutually unbiased bases are a primitive of
quantum theory, as they embody the importance
of the superposition principle. In this paper
we have used them to develop a comprehensive
quantum theory of the phases as complementary
to inversions in finite-dimensional systems.

The construction presented in this paper is
devoid of any ambiguity associated with the
non-uniqueness of polar decomposition of
ladder operators. In prime dimensions, phase
operators and inversions are elegantly related
by a finite Fourier transform, much like
positions and momenta are related by an
ordinary Fourier transform in infinite-dimensional
systems, and provides an appealing way of
treating a concept as central as the phases
of a system.

\section*{Acknowledgments}

We would like to acknowledge stimulating
discussions with Prof. Gunnar Bj\"ork.
The work of Hubert de Guise is supported by
NSERC of Canada.

\end{document}